\documentclass[journal]{IEEEtran}
\usepackage{lineno,hyperref}
\modulolinenumbers[5]
\usepackage[table,xcdraw]{xcolor}
\usepackage{xcolor}
\usepackage{amssymb}
\usepackage{ifpdf}
\usepackage{enumitem}
\usepackage{booktabs}
\usepackage{float}  
\usepackage[cmex10]{amsmath}
\usepackage{amsfonts}
\usepackage{graphicx}
\usepackage{multirow}
\usepackage{algorithm}
\usepackage{physics}
\usepackage{adjustbox}
\usepackage{amsthm}

\usepackage[algo2e]{algorithm2e}
\SetArgSty{textnormal}
\usepackage{setspace}
\usepackage{enumitem}
\usepackage{subfig} 
\usepackage{array}
\usepackage{stfloats}
\usepackage{url}
\usepackage[countmax]{subfloat}
\usepackage{xcolor}
\usepackage{fixmath}
\usepackage{epstopdf}
\usepackage{balance}
\begin{document}
\title{Consensus Tracking of an Underwater Vehicle Using Weighted Harmonic Mean Density}
\author{Ved Prakash Dubey
        and~Shovan Bhaumik
\thanks{Ved Prakash Dubey and Shovan Bhaumik are with the Department of Electrical Engineering, Indian Institute of Technology Patna (ved$\_$1921ee10@iitp.ac.in; shovan.bhaumik@iitp.ac.in)} }
\maketitle
\begin{abstract}
This paper addresses an underwater target tracking problem in which a large number of sonobuoy sensors are deployed on a surveillance region. The region is divided into several sub-regions, where a single tracker, capable of generating track is installed. Each sonobuoy can measure the direction of arrival of acoustic signals (known as bearing angles) and communicate the measurements with the local tracker. Further, each local tracker can communicate with all other trackers, where each of them can exchange their estimate and finally a consensus is reached. We propose a weighted harmonic mean density (HMD) based tracking to reach a consensus and provide a solution for the fusion of Gaussian densities. In this approach, optimal weights are assigned by minimizing the Kullback-Leibler divergence measure. Performance of the proposed method is measured using root mean square error, percentage of track divergence, and normalized estimation error squared. Simulation results demonstrate that the optimized HMD-based fusion outperforms existing fusion methods during a distributed tracking.
\end{abstract}
\IEEEpeerreviewmaketitle

\section{Introduction}
\label{sec:introduction}
In the last few decades, there have been substantial advances in communication and sensor technology, leading to the increased adoption of multi-sensor systems in the field of underwater tracking \cite{isbitiren2011three, ghafoor2019overview, hou2022bearing, zhao2024novel}. In a single-sensor system, the received information becomes highly corrupted or lost for a long distance sensing \cite{ macaulay2020field, das2025constrained}. Therefore, sensors should be placed moderately close to the signal source to obtain more precise information. To address this issue, a multi-sensor tracking system was developed for underwater surveillance and tracking applications \cite{majumder2001multisensor, zhang2012distributed, taghavi2016multisensor, kim2025tracking}.

The tracking process using multiple sensors can be categorized into three groups: centralized, decentralized, and distributed tracking \cite{wei2019distributed, li2015weighted}. Among these, decentralized and distributed tracking are popular for their robustness in handling sensor failures and accommodating addition or removal of sensors and trackers \cite{jiao2021deep, wei2019distributed}. Distributed tracking over a sensor network involves two types of nodes: sensor nodes and processing nodes. Sensor nodes possess only sensing and communication capabilities, allowing them to sense a physical quantity from the environment and also be able to send their measured information to the processing nodes. On the other hand, processing nodes are equipped with both computational and communication capabilities. They can process the information received from sensors to estimate the system states and subsequently share the estimated information with neighboring processing nodes \cite{wei2019distributed, deng2015multisensor}. In tracking applications, the processing nodes are also called trackers. 

Using prior knowledge of the target, each tracker identifies sensors located closer to the target that have more accurate and informative measurements \cite{tharmarasa2007large}. The selected sensor then generates measurement samples of the suspected target and forwards them to their respective trackers. Each tracker utilizes only the measured information of its local sensors to determine the state of the target. However, it is well known that sensors located closer to the target provide measurements with greater variation in their bearing information. Consequently, only the tracker situated in the sub-region containing the target provides an accurate estimate. Therefore, it is necessary to fuse the estimated information across all trackers to achieve a consensus, ensuring that the state estimates of all trackers are accurate and consistent.

To reach a consensus \cite{li2015weighted} among the trackers, they share the probability density function (pdf) (also called tracks) with their neighboring trackers. Track fusion takes place at each tracker to reach consensus between their track densities. 
In the context of information shared between the tracking nodes, the consensus filters are broadly classified into three categories: consensus on the estimate (CE), consensus on measurement (CM), and consensus on information (CI) \cite{li2003survey, li2015weighted}. These filters are restricted to the Bayesian framework and have limited effectiveness. To overcome this limitation, alternative consensus strategies are employed to merge density information from multiple trackers. These techniques are commonly known as track-to-track fusion, as they consolidate the probability density functions \cite{tian2009track}.

To reach a consensus, computation of mean plays a pivotal role. Three classical approaches for computing the mean of a data set are the arithmetic, geometric, and harmonic means. Among these, the arithmetic and geometric means together with their weighted generalizations have been extensively studied in the context of track density fusion and data association \cite{nielsen2019jensen}. Among them, arithmetic density fusion is known to overestimate uncertainty, thereby degrading the estimation accuracy \cite{niculescu2006convex}. On the other hand, geometric mean density fusion does not admit a closed-form solution because of the non-integral exponent of the density function \cite{hurley2002information, mahler2000optimal}. To address this, weighted inverse covariance fusion, also known as Chernoff fusion, is proposed in the literature as an approximation of weighted geometric fusion \cite{wei2019distributed,uney2011information}. The harmonic mean density is free from both of these limitations.  

In \cite{sharma2024pooling}, a weighted harmonic mean fusion method was proposed to combine two density functions. To fuse multiple densities, a sequential fusion strategy was suggested in \cite{sharma2024pooling}. This approach is computationally not efficient and requires successive pairwise fusions with an optimization step at each stage to determine the optimal weight. To address this limitation, here we introduce a method to compute the harmonic mean density among multiple pdfs in a single step. The weights of the densities are found by minimizing Kullback-Leibler divergence measure.   As in the proposed framework, the weights corresponding to all densities are obtained through a single optimization step, thereby yielding a more efficient and scalable solution for multi-track fusion.

The effectiveness of the proposed fusion method is evaluated by comparing its tracking performance with other fusion techniques in a distributed tracking environment. The comparison is conducted in terms of root mean square error (RMSE) and percentage of track divergence, while consistency is verified using the average normalized estimation error squared (ANEES). In simulation, the proposed harmonic mean density fusion demonstrates robust performance when deterministic sample point filters and the shifted Rayleigh filter are adopted to generate tracks.

\section{Tracking Problem in a Surveillance Region} \label{Problem formulation}
As discussed earlier, we consider a scenario in which a large number of battery-powered sonobuoys are airdropped and uniformly distributed across the sea surface within the surveillance region. Each sensor is capable of measuring the direction of arrival of an acoustic signal emitted by an underwater target moving within the region. Unlike centralized approaches (e.g., \cite{tharmarasa2007large}), we adopt a distributed architecture in which multiple trackers are uniformly deployed, thereby partitioning the surveillance region into several sub-regions. All sensors are synchronized with their nearest tracker and transmit their measurements to it. Each tracker then estimates the target state locally using the measurements from its associated sensors.

\begin{figure}[htbp] 
	\centering
	\includegraphics[width=0.48\textwidth]{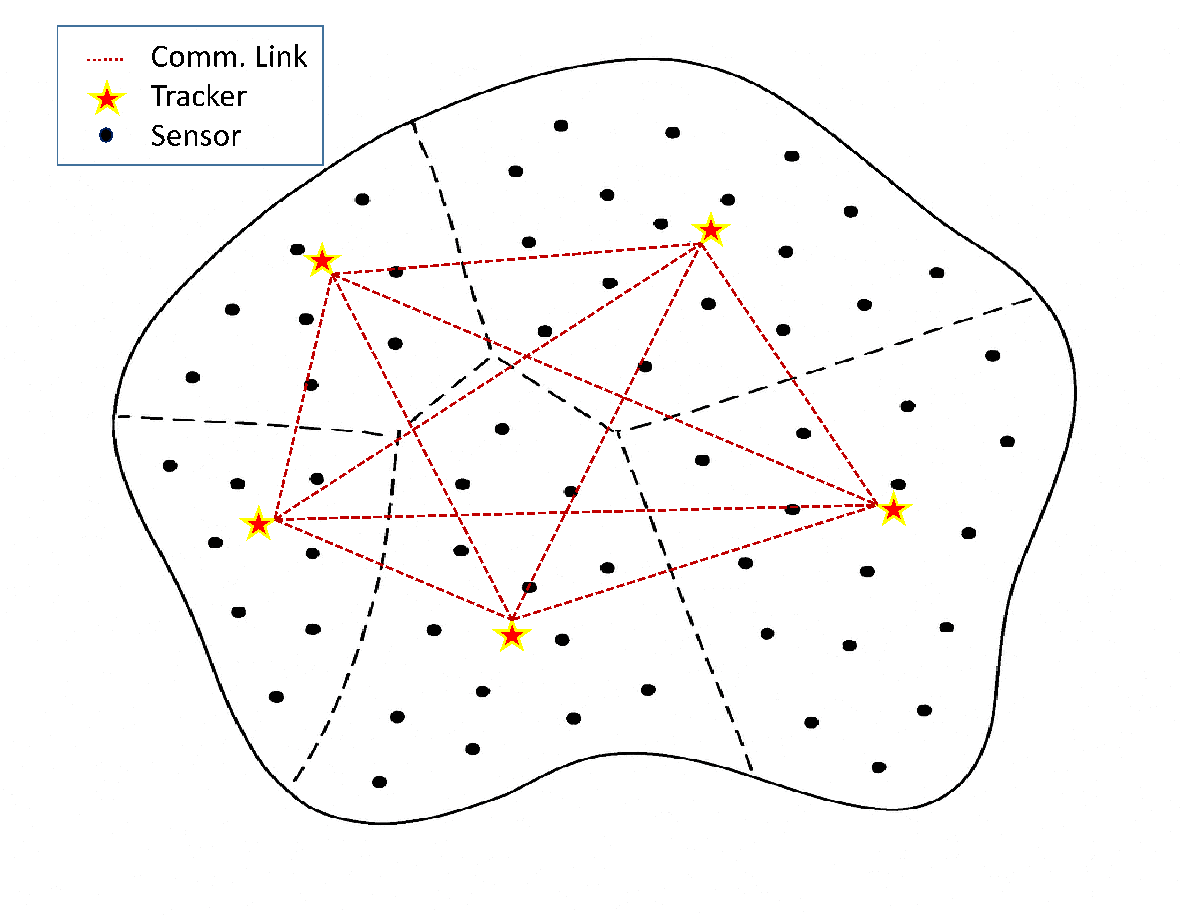} 
	\caption{Schematic representation of a distributed tracking architecture  where the sensors are deployed uniformly. Here, the whole area is divided in several sub-regions, and trackers are placed inside each sub-region and they are connected with each other through a communication link.}
	\label{fig:schematic} 
\end{figure}

This distributed framework is inherently more robust, as the failure of a communication channel or an individual tracker has minimal impact on the overall tracking performance. Moreover, it offers reduced processing time compared to centralized tracking. An example of the distributed tracking architecture is illustrated in Fig.~\ref{fig:schematic}. In this figure, we can see the distribution of sensors and trackers in arbitrary surveillance area.  The trackers are  interconnected via wireless communication channels, allowing them to exchange their estimated information.  The key assumptions underlying the considered scenario are summarized as follows.
\begin{itemize}
	\item The sensors are stationary (unaffected by sea currents) and they are uniformly distributed throughout the surveillance region.
	\item The total area is subdivided in several regions, and a tracker is placed in each subregion where each sensor is linked to its nearest tracker.
	\item To prolong battery life, only a few sensors are activated during a specific interval to transmit their measurements.
	\item All sensors are identical, passive devices capable of measuring the bearing angle of an acoustic source.
	\item Each tracker performs local tracking using the available measurements and it generates a track.
	\item Each tracker communicates with its neighboring trackers, sharing its estimated track. Through track fusion, each node produces a consensus tracking result.
\end{itemize}

\subsection{Process model} \label{process_model}
\subsubsection*{Scenario 1: constant velocity model }
In this scenario, the target follows a straight-line motion and it is modeled as a nearly constant velocity \cite{bar2004estimation}. The process dynamics is defined by
\begin{equation} \label{process1}  \mathcal{X}_{k} = F\mathcal{X}_{k-1} +\eta_{k},   
\end{equation} 
where $ \mathcal{X}_k,\eta_k \in \mathbb{R}^{n_{\mathcal{X}}}$,  $\mathcal{X}_{k} = \begin{bmatrix} x_{k} & y_{k}  & \dot{x}_{k} & \dot{y}_{k}  \end{bmatrix}^\top$ is a target state vector. $x_k$, $y_k$  are positions along $X$ and $Y$ axis, respectively at time step $k$. $F$ is the system matrix which can be expressed as  $F =\begin{bmatrix} I_2 & T I_2 \\ 0_2 & I_2  \end{bmatrix}$, where $I_2$ is the $2\times2$ identity matrix, $0_2$ the $2\times2$ zero matrix, and $T$ is the sampling interval. $\eta_{k}$ is the process noise which is assumed to be white, Gaussian with zero mean and covariance $Q_k$ \emph{i.e.} $\eta_{k} \sim \mathcal{N} (0, Q_k)$. The expression of $Q_k$ is 
\begin{equation}
Q_k= {q}_1\begin{bmatrix} \frac{T^3}{3} I_2 &  \frac{T^2 }{2} I_2 \\ \frac{T^2}{2}I_2  & TI_2  \end{bmatrix},
\end{equation}
where  ${q}_1$ is the intensity of  process noise.

\subsubsection*{Scenario 2: constant turn rate model}
In this scenario, the target motion is along a coordinated turn (CT) and is modeled in the $X-Y$ plane with a nearly constant turn rate \cite{li2003survey}. The process dynamics are defined by
\begin{equation} \label{process2} 
\mathcal{X}_{k} = F_{k-1}\mathcal{X}_{k-1} +\eta_{k},    
\end{equation}
where $\mathcal{X}_{k} = \begin{bmatrix} x_{k}  & y_{k} & \dot{x}_{k} & \dot{y}_{k} & \Omega_k\end{bmatrix}^\top$, $\Omega_k$ is the angular turn rate of the target, positive for counterclockwise rotation and negative for clockwise rotation, and the expression of $F_k$ is given by

\begin{equation} \label{system2}  
F_k = \begin{bmatrix}
1 & 0 & \frac{\sin\Omega_k T}{\Omega_k} & -\frac{1-\cos\Omega_k T}{\Omega_k} & 0  \\
0 & 1 & \frac{1-\cos\Omega_k T}{\Omega_k} & \frac{\sin\Omega_k T}{\Omega_k} & 0 \\
0 & 0 & \cos\Omega_k T & -\sin\Omega_k T & 0 \\
0 & 0 & \sin \Omega_k T & \cos\Omega_k T & 0 \\
0 & 0 & 0 & 0 & 1
\end{bmatrix}.            
\end{equation}
The process noise covariance matrix $Q_k$  corresponding to this system is given by $Q_k= \operatorname{diag}([{q}_1Q_{1,k}, ~ {q}_2T])$, where
\begin{equation}
Q_{1,k}= \begin{bmatrix} 
\frac{2s_{\Omega_k}}{\Omega_k^3} & 0 & \frac{c_{\Omega_k}}{\Omega_k^2}  & -\frac{s_{\Omega_k}}{\Omega_k^2} \\
0 & \frac{2s_{\Omega_k}}{\Omega_k^3} & \frac{s_{\Omega_k}}{\Omega_k^2}  & \frac{c_{\Omega_k}}{\Omega_k^2} \\
\frac{c_{\Omega_k}}{\Omega_k^2}  & \frac{s_{\Omega_k}}{\Omega_k^2} & T  & 0 \\
-\frac{s_{\Omega_k}}{\Omega_k^2} & \frac{c_{\Omega_k}}{\Omega_k^2} & 0  & T
\end{bmatrix},
\end{equation}
where $s_{\Omega_{k}}=\Omega_{k} T - \sin \Omega_{k} T$, $c_{\Omega_{k}}=1-\cos \Omega_{k} T$ and ${q}_2$ is the process noise intensity corresponding to its turn rate.

\subsection{Measurement model}
The sensors are deployed in the surveillance area and they are capable to measure the bearing angle of the target at their location. The mathematical expression representing the measurement acquired by the $i$th sensor in the $j$th sub-region at the $k$th instant is denoted by  
\begin{equation}
Z_{i,k}^j=\theta_{i,k}^j + \upsilon_{i,k}^j,
\end{equation}
where
\begin{equation}
\theta_{i,k}^j= h(\mathcal{X}_k, \mathcal{X}_i^{s,j}) = \tan^{-1}\left(\dfrac{x_k - x_i^{s,j}}{y_k-y_i^{s,j}}\right),     
\end{equation}
where $ (x_i^{s,j}, y_i^{s,j}) $ is the $i$th sensor coordinate in $j$th subregion, $\theta_{i,k}$ is the bearing angle of the sensor, $\tan^{-1}(\cdot)$ denotes the four-quadrant inverse tangent function, and the angles $\{Z_{k,i}^j, \theta_{i,k}^j, \upsilon_{i,k}^j\} \in (-\pi, \pi]$. The term $ \upsilon_{i,k}$ represents the measurement noise, assumed to be white, Gaussian with zero mean and covariance $\mathbb{E} [\upsilon_{i,k}^2]= R_{i,k}$.

At a particular time instant, a local tracker selects a subset of sensors (at least two) from the available set and receives their measurements. The selection of sensors for transmitting their measurements is carried out by minimizing the trace of the Fisher information matrix (FIM). For a detailed discussion on the optimization procedure, readers are referred to \cite{dubey2023tracking}. Thus, at any instant $k$, the measurements received by the $j$th local tracker are given by 
\begin{equation}
\mathcal{Y}_k^j =
\{Z_{i,k}^j \; | \; \mathcal{S}_{i,k}^j = 1\},
\end{equation}
where $\mathcal{S}_{i,k}^j \in \{0,1\}$ denotes the status of the $i$th sensor for the $j$th tracker at time $k$. Specifically, $\mathcal{S}_{i,k}^j = 1$ if the $j$th tracker selects the measurement from the $i$th sensor, and $\mathcal{S}_{i,k}^j = 0$ otherwise.


In the described scenario, when a target is suspected within the surveillance area, all distributed trackers activate a subset of sensors based on the detected target location. Each tracker then estimates the target’s track using the measurements from its activated sensors. These tracks are generated locally, and their estimation accuracy depends on the proximity of the target to the activated sensors; hence, trackers closer to the target provide better estimates than those farther away. Each tracker shares its estimated track density with its neighboring trackers, enabling them to fuse the densities and reach a consensus estimate.  

The objective is to find a consensus density function so that we can get accurate estimation results with minimal error. Our approach employs optimal weighted mean harmonic density fusion, where the weights are determined by formulating a cost function based on the symmetrized Kullback–Leibler divergence measure.

\section{Harmonic Mean Density Fusion} \label{density_fusion_hmd}
As stated above, the local tracker estimates the target state using the measurements of the sensors located in the corresponding subregion. Let us represent the estimated pdf of $j$th tracker as $p_j(\mathcal{X}_k|\mathcal{Y}_k^j)$ or briefly by $p_j(\mathcal{X}_k)$. Our objective is to fuse all such pdfs and obtain an overall consensus pdf of the target states. The generalized weighted mean density $p_\omega(\mathcal{X}_k)$ for mixing pdfs obtained from $N$ number of trackers \emph{viz.}  $ p_1(\mathcal{X}_k), \cdots , p_{N}(\mathcal{X}_k) $ can represented as \cite{uney2011information}
\begin{equation}
p_{\omega}(\mathcal{X}_k) = \dfrac{\mathcal{M}_{\omega}( p_1(\mathcal{X}_k), \ldots ,  p_{N}(\mathcal{X}_k))}{\zeta_{\mathcal{M}_\omega}(p_1(\mathcal{X}_k), \ldots ,  p_N(\mathcal{X}_k))}, \label{mdf}
\end{equation}
where $ \mathcal{M}_{\omega}(.) $ is the weighted mean function, $\omega$ is the weight vector, and the normalization constant $\zeta_{\mathcal{M}_{\omega}}(.)$ ensures fusion density to a valid probability density function.
The fused density can be obtained using arithmetic mean, geometric mean, or harmonic mean formulations. In this section, we derive the harmonic mean density (HMD) function for the general case of combining an arbitrary number of densities. For the fusion process, we adopt a one-step batch fusion of multiple pdfs, which significantly reduces computation time for both weight optimization and density fusion compared to the recursive procedure suggested in \cite{sharma2024pooling}.
\subsection{Arithmetic and geometric mean density}

The solution of the mixing probabilities in Eqn.~(\ref{mdf}) for arithmetic mean density (AMD) fusion is represented as a convex combination of probability densities \cite{niculescu2006convex}. In the case of geometric mean density (GMD) fusion, the fused density is obtained as an exponential mixture of the individual densities \cite{hurley2002information}. 
These fused densities can be expressed as:
\begin{equation}
\begin{split}
\mathcal{M}_{\omega}^a(\mathcal{X}_k) 
& = \omega_1 p_1(\mathcal{X}_k) + \cdots + \omega_N p_N(\mathcal{X}_k), \\ &= \sum_{j=1}^N \omega_j p_j(\mathcal{X}_k), 
\end{split}
\end{equation}

\begin{equation} \label{gmdfeq} 
\begin{split}
\mathcal{M}_{\omega}^g(\mathcal{X}_k) 
&= p_1^{\omega_1}(\mathcal{X}_k) \times \cdots \times p_N^{\omega_N}(\mathcal{X}_k),\\ 
& = \prod_{j =1} ^Np_j^{\omega_j}(\mathcal{X}_k),
\end{split}
\end{equation}
where weights are $ \{\omega_1, \cdots, \omega_N\} \in [0, 1]$, and $ \displaystyle \sum_{j = 1}^N \omega _j  = 1 $.  The weight represents the contribution of each tracker's density in the fused density.
\subsection{Harmonic mean density function}
Previous studies \cite{niculescu2006convex, hurley2002information, mahler2000optimal} have explored the use of AMD and GMD functions for probabilistic fusion. However, both approaches exhibit notable limitations. AMD fusion tends to overestimate uncertainty, which can compromise the consistency and informativeness of the resulting estimates. GMD fusion, while theoretically appealing, lacks a closed-form solution due to the non-integral exponent in its density formulation. To address this, Chernoff fusion-also known as exponential mixture density fusion, is often employed as an approximation to GMD, typically by matching the first two moments with a Gaussian surrogate \cite{nielsen2022revisiting}. On the other hand, harmonic mean density (HMD) fusion avoids these problems, offering a closed-form, consistent, and reliable framework for producing a consensus probability density function.

The mixture of weighted HMD fusion mixing probabilities provided in Eqn.~(\ref{mdf}) is given by \cite{sharma2024pooling}
and is defined by
\begin{equation}\label{hmdf}
\begin{split} \mathcal{M}_{\omega}^h(\mathcal{X}_k) 
&= \Big(\frac{\omega_j}{p_j(\mathcal{X}_k)} + \cdots + \frac{\omega_j}{p_j(\mathcal{X}_k)} \Big)^{-1},\\
&= \Big( \sum_{j=1}^{N} \frac{\omega_j}{p_j(\mathcal{X}_k)} \Big)^{-1}.
\end{split}
\end{equation}
The harmonic pdf is proportional to its mixing probability and simplified expression is represented by
\begin{equation}\label{hmdfeq}
\begin{split}
p_{\omega}^h(\mathcal{X}_k) \propto \mathcal{M}_{\omega}^h(\mathcal{X}_k) 
& = \frac{\displaystyle \prod_{j=1}^{N} p_j(\mathcal{X}_k)}{\displaystyle \sum_{j =1}^{N} \omega_j \prod_{i=1,\, i \neq j}^{N} p_i(\mathcal{X}_k)},
\end{split}
\end{equation}
where $\omega_j$ is the weight associated with the $j$-th component of the distribution. Let $\{p_j(\mathcal{X}_k)\}_{j=1}^N$ denote $N$ multivariate Gaussian densities, where $ p_j(\mathcal{X}_k) = \mathcal{N}(\hat{\mathcal{X}}_{k|k}^j,\, P_{k|k}^j $). Substituting into \eqref{hmdfeq}, the weighted HMD mixture can be expressed as
\begin{equation}
\begin{split}
p_{\omega}^h(\mathcal{X}_k)
\propto & \dfrac{\displaystyle \prod_{j=1}^{N} \mathcal{N}(\hat{\mathcal{X}}_{k|k}^j, P_{k|k}^j)}{\displaystyle \sum_{j=1}^{N} \omega_j \prod_{i=1,\, i \neq j}^{N} \mathcal{N}(\hat{\mathcal{X}}_{k|k}^i, P_{k|k}^i)} \\
= & \dfrac{\displaystyle \prod_{j=1}^{N} \mathcal{N}(\hat{\mathcal{X}}_{k|k}^j, P_{k|k}^j)}{\displaystyle \sum_{j=1}^{N} \omega_j \,\mathcal{N}(\hat{\mathcal{X}}_{k|k}^{c,j}, P_{k|k}^{c,j})}, 
\end{split}
\end{equation}
where each term in the denominator consists of the weight $\omega_j$ multiplied by the product of $N-1$ Gaussian densities. If we assume that each tracker approximates the estimated density function with a Gaussian distribution, then their multiplication will also remain Gaussian. So, we write $  \prod_{i=1,\, i \neq j}^{N}  \mathcal{N}(\hat{\mathcal{X}}_{k|k}^i, P_{k|k}^i)=\mathcal{N}(\hat{\mathcal{X}}_{k|k}^{c,j}, P_{k|k}^{c,j})$, where $c$ is used to denote the combined Gaussian parameters. The mean and covariance of the combined Gaussian are given by \cite{kang2020data}
\begin{equation}
P_{k|k}^{c,j} = \Big[ \sum_{i=1,\, i\neq j}^N {P_{k|k}^{i}}^{-1} \Big]^{-1}, 
\end{equation}
and
\begin{equation}
\hat{\mathcal{X}}_{k|k}^{c,j} = P_{k|k}^{c,j} \Big[ \sum_{i=1,\, i \neq j}^N {P_{k|k}^{i}}^{-1} \hat{\mathcal{X}}_{k|k}^i \Big].
\end{equation}

The denominator of \eqref{hmdfeq} is a weighted sum of Gaussian densities, which is non-Gaussian in general. It can, however, be approximated by a single Gaussian using moment matching, similar to the weighted AMD fusion \cite{niculescu2006convex}. Specifically,
\begin{equation}
\sum_{j=1}^{N} \omega_j \,\mathcal{N}\!\big(\hat{\mathcal{X}}_{k|k}^{c,j}, P_{k|k}^{c,j}\big) 
\;\approx\; 
\mathcal{N}\!\big(\hat{\mathcal{X}}_{k|k}^{eq}, P_{k|k}^{eq}\big),
\end{equation}
where $\hat{\mathcal{X}}_{k|k}^{eq}$ and $P_{k|k}^{eq}$ denote the mean and covariance of the approximated Gaussian, and are given by
\begin{equation}
\hat{\mathcal{X}}_{k|k}^{eq} = \sum_{j=1}^{N} \omega_j \hat{\mathcal{X}}_{k|k}^{c,j},
\end{equation}
and
\begin{equation}
P_{k|k}^{eq} = \sum_{j=1}^{N} \omega_j \big[ P_{k|k}^{c,j} 
+ (\hat{\mathcal{X}}_{k|k}^{eq} - \hat{\mathcal{X}}_{k|k}^{c,j})(\hat{\mathcal{X}}_{k|k}^{eq} - \hat{\mathcal{X}}_{k|k}^{c,j})^\top \big].  
\end{equation}
Now, from \eqref{hmdfeq} using the Gaussian approximation, the fused density becomes
\begin{equation}
p_{\omega}^h(\mathcal{X}_k) 
\;\approx\; 
\dfrac{\displaystyle \prod_{j=1}^{N} \mathcal{N}(\hat{\mathcal{X}}_{k|k}^j, P_{k|k}^j)}{\mathcal{N}(\hat{\mathcal{X}}_{k|k}^{eq}, P_{k|k}^{eq})}
= \mathcal{N}(\hat{\mathcal{X}}_{k|k}^f, P_{k|k}^f),
\end{equation}
where the fused mean and covariance are obtained as \cite{kang2020data}
\begin{equation}
P_{k|k}^f = \Big[  \sum_{j=1}^{N} {P_{k|k}^j }^{-1} -{ P_{k|k}^{eq}}^{-1} \Big]^{-1},
\end{equation}
\begin{equation}
\hat{\mathcal{X}}_{k|k}^f = P_{k|k}^f \Big[  \sum_{j=1}^{N} {P_{k|k}^j}^{-1} \hat{\mathcal{X}}_{k|k}^j  - {P_{k|k}^{eq}}^{-1} \hat{\mathcal{X}}_{k|k}^{eq} \Big].  
\end{equation}

\subsection{Weight optimization}
In density fusion, it is crucial to assign suitable weights to the individual distributions to obtain a probability density function of states. These weights are typically derived by formulating a cost function, where a symmetrical divergence measure serves as the performance criterion for evaluating distribution \cite{uney2011information}. An objective function for determining the weights of individual densities is constructed to minimize the dispersion of the divergences between the fused density $p_{\omega}(\mathcal{X}_k)$ and the individual densities $\{p_i(\mathcal{X}_k)\}_{i=1}^N$. We formulate an optimization problem as described below. 
\begin{equation}
\omega^*_k =  \arg\min_{\omega} \sum_{i=1}^N
\Big(\mathcal{D}\big(p_{\omega}(\mathcal{X}_k) : p_i(\mathcal{X}_k)\big) - \overline{\mathcal{D}}\Big)^2,
\end{equation}
subject to
\begin{equation}
0\leq \omega_i \leq 1,
\end{equation}
and 
\begin{equation}
\sum_{i=1}^N \omega_i = 1,
\end{equation}
where $\mathcal{D}$ denotes a symmetric divergence measure and 
\begin{equation}
\overline{\mathcal{D}} = \frac{1}{N}\sum_{i=1}^N \mathcal{D} \big(p_{\omega}(\mathcal{X}_k) : p_i(\mathcal{X}_k)\big).
\end{equation}
In this work we have consider Kullback–Leibler (KL) divergence as a measure of diveregence. Since the KL divergence is inherently asymmetric, it must be symmetrized for use in this context. The symmetrized KL divergence between two probability densities $p_i(\mathcal{X}_k)$ and $p_j(\mathcal{X}_k)$ is given by
\begin{equation}
\begin{split}
\mathcal{D}\big(p_i(\mathcal{X}_k):p_j(\mathcal{X}_k)\big)
&= \tfrac{1}{2}\big[
\mathcal{D}_{\mathrm{KL}}\big(p_i(\mathcal{X}_k):p_j(\mathcal{X}_k)\big) \\
&\quad
+\,\mathcal{D}_{\mathrm{KL}}\big(p_j(\mathcal{X}_k):p_i(\mathcal{X}_k)\big)
\big],
\end{split}
\end{equation}
where the KL divergence between them is defined as \cite{zhang2023properties}:
\begin{equation}
\mathcal{D}_{\mathrm{KL}}(p_i(\mathcal{X}_k) : p_j(\mathcal{X}_k)) = \int p_i(\mathcal{X}_k) \ln  \frac{p_i(\mathcal{X}_k)}{p_j(\mathcal{X}_k)} d\mathcal{X}_k.
\end{equation}
In the case of $p_i(\mathcal{X}_k) \sim \mathcal{N}(\hat{\mathcal{X}}_{k|k}^i,\;P_{k|k}^i)$ and 
$p_j(\mathcal{X}_k) \sim \mathcal{N}(\hat{\mathcal{X}}_{k|k}^j,\;P_{k|k}^j)$, the KL divergence is expressed by
\begin{equation} \label{KL_for_Gaussian_new}
\begin{split}
\mathcal{D}_{\mathrm{KL}}(\cdot)
&= \tfrac{1}{2} \Big[
\mathrm{tr}\!\big({P_{k|k}^j}^{-1} P_{k|k}^i\big)
- n_{\mathcal{X}}
+ \ln\!\frac{|P_{k|k}^j|}{|P_{k|k}^i|}
\Big] \\
&\quad
+ \tfrac{1}{2}
(\hat{\mathcal{X}}_{k|k}^i - \hat{\mathcal{X}}_{k|k}^j)^\top
{P_{k|k}^j}^{-1}
(\hat{\mathcal{X}}_{k|k}^i - \hat{\mathcal{X}}_{k|k}^j),
\end{split}
\end{equation}
where $n_\mathcal{X}$ is the dimensionality of the density function, $|\cdot|$ denotes the matrix determinant, $\mathrm{tr}(\cdot)$ denotes the trace of a matrix. A derivation of the above expression is provided in \hyperref[KL_for_Gaussian]{Appendix~\ref*{KL_for_Gaussian}}.

\section{Simulation Results}
To simulate the designed problem, we considered a total of \(100\) sonobuoys distributed over a $(10 \times 10) \, \text{km}^2$ surveillance region. A uniform distribution is adopted to realize the sensor placement. The entire area was divided into four $(5 \times 5) \, \text{km}^2$ square sub-regions, and four trackers $(N=4)$ were deployed, one assigned to each sub-region. All trackers were positioned near the center of their respective sub-regions, ensuring that all sensors could transmit their measurements while minimizing energy expenditure throughout the tracking interval.  


Initially, all sensors were assumed to be in passive mode. Whenever a target was suspected within the surveillance region, each tracker activates two sensors from its sub-region, and allowing them to transmit their measurements. These sensors are strategically chosen by minimizing a cost function derived from their Fisher information matrix, as outlined in \cite{dubey2021selection, tharmarasa2007large}. Using the bearing angle of these chosen sensors, the designed tracking system becomes observable, and the tracker could triangulate the target position within its estimation algorithm. 

Two tracking scenarios are considered: a non-maneuvering target modeled as nearly constant velocity (CV) and a maneuvering target modeled as nearly constant turn (CT). At any given time, only a single target is assumed to be present for tracking. Both the scenarios as described in the section \ref{process_model} are simulated for 36 minutes with a sampling interval of $T = 0.25$~min, and the remaining simulation parameters are listed in Table~\ref{tab_simulation_parameters}. 

\begin{figure}[htbp]
	\centering
		\centering
		\includegraphics[width=\linewidth]{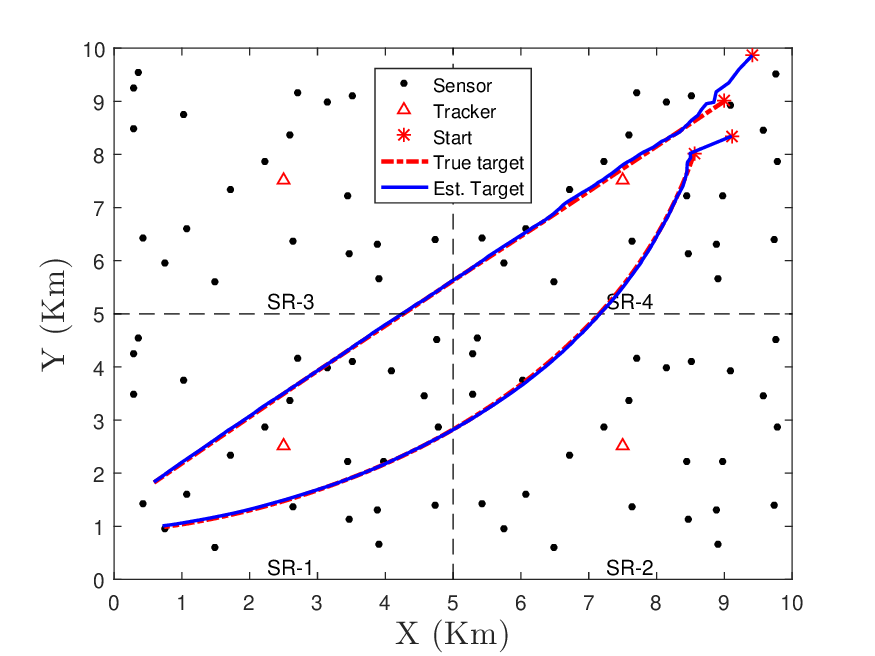}
		\caption{Illustration of the tracking scenarios with sensor locations, true and estimated target trajectory, when the target moves with nearly constant velocity and with a constant turn rate.}
		\label{fig:scenario}
\end{figure}

The tracking scenarios are illustrated in Fig.~\ref{fig:scenario}, where black dots represent sensors' location, and red triangles denote tracker positions. The surveillance region is divided into sub-regions (outlined by dotted lines), each labeled accordingly. The true target and its estimated initial position are marked with red stars, while their respective trajectories are shown by dotted red and blue lines. Furthermore, all trackers were assumed to mutually interconnected, sharing their tracking information with one another. In this case, fusion was performed at each node, and a single-step fusion was sufficient to achieve consensus, resulting in identical track densities across all trackers after fusion.  
\begin{table}[htbp]
	\centering
	\caption{Tracking parameters}
	\label{tab_simulation_parameters}
	\renewcommand{\arraystretch}{1.2}
	\begin{tabular}{p{3.8cm}p{1.5cm}p{1.5cm}}
		\hline\hline
		Parameters & Scenario~1 & Scenario~2 \\
		\hline
		No. of operational sensors ($n_s$) & 100 & 100 \\
		No. of selected sensors ($n$) & 2 & 2 \\
		Sampling interval ($T$) & 0.25~min & 0.25~min \\
		Sensor selection interval & 2~min & 2~min \\
		Process noise intensities ($q_1, q_2$) & $1.944~\text{m}^2/\text{min}^3$ & $0.01~\text{rad}^2/\text{min}^3$ \\
		Measurement noise ($R_{i,k}$) & $(2^\circ)^2$ & $(2^\circ)^2$ \\
		Initial target location & $(9,9)$~km & $(8.5,8)$~km \\
		Target speed ($v^t$) & 10~knots & 10~knots \\
		Target initial course angle & $-130^\circ$ & $-165^\circ$ \\
		Target turn rate ($\Omega$) & -- & $-1.84^\circ/\text{min}$ \\
		\hline\hline
	\end{tabular}
\end{table}

\subsection{Initialization}
Prior to estimation, all tracks are initialized using the first two measurements from two selected sensors, following the method described in~\cite{dubey2021selection}. The target location and corresponding uncertainty can be obtained by triangulating the bearing measurements from any two sensors. Let us consider the two sensors,  positioned at $(x_1^s, y_1^s)$ and $(x_2^s, y_2^s)$, and their respective bearing measurements at time $k$ are $Z_{1,k}$ and $Z_{2,k}$. The intersection of bearing lines gives the target position and is obtained by
\begin{equation}
\begin{split}
x_k &= x_1^s +  \lambda \sin(Z_{1,k}),\\
y_k &= y_1^s + \lambda\cos(Z_{1,k}),
\end{split}
\end{equation}
where $ \lambda = \dfrac{(y_2^s-y_1^s)\sin(Z_{2,k}) - (x_2^s-x_1^s)\cos(Z_{2,k})}{\sin(Z_{2,k}-Z_{1,k})} $. Here, $Z_{2,k} \neq Z_{1,k}$, and the difference between the two measurements is sufficiently large to enable reliable and more accurate triangulation.  Using two consecutive triangulated position estimates, the target state is initialized by
\begin{equation}
\hat{\mathcal{X}}_{0|0} = 
\begin{bmatrix}
x_0,  ~ y_0, ~ \dfrac{x_1 - x_0}{T}, ~ \dfrac{y_1 - y_0}{T}
\end{bmatrix}^\top ,
\end{equation}
where $(x_0,y_0)$ and $(x_1,y_1)$ are successive triangulated positions. The corresponding initial covariance is
\begin{equation}
P_{0|0} = \operatorname{diag}([ P_{xy}, ~ \tfrac{v_m^2}{3} I_2 ]) ,
\end{equation}
with $P_{xy}$ denoting the position uncertainty obtained by first-order propagation of bearing measurement variance and determined by
\begin{equation}
P_{xy} = J R J^\top,
\end{equation}
where $J = \nabla_{\mathcal{Y}_0}(H\hat{\mathcal{X}}_{0|0})$ denotes the Jacobian of $H\hat{\mathcal{X}}_{0|0}$ with respect to the measurement vector $\mathcal{Y}_0$, $H = [I_2~0_2]$ is the transformation matrix, $\mathcal{Y}_0 = [Z_{1,0},~ Z_{2,0}]^\top$ is the combined sensor measurement, and $R = \operatorname{diag}([R_{1,0},~ R_{2,0}])$ is the corresponding covariance. Assuming the target speed along each axis is uniformly distributed in $[-v_m,~v_m]$, the variance of each velocity component is $\tfrac{v_m^2}{3}$, defining the velocity block of $P_{0|0}$.

\begin{figure*}[htbp]
	\centering
	\begin{minipage}[b]{0.48\textwidth}
	\centering
	\includegraphics[width=\linewidth]{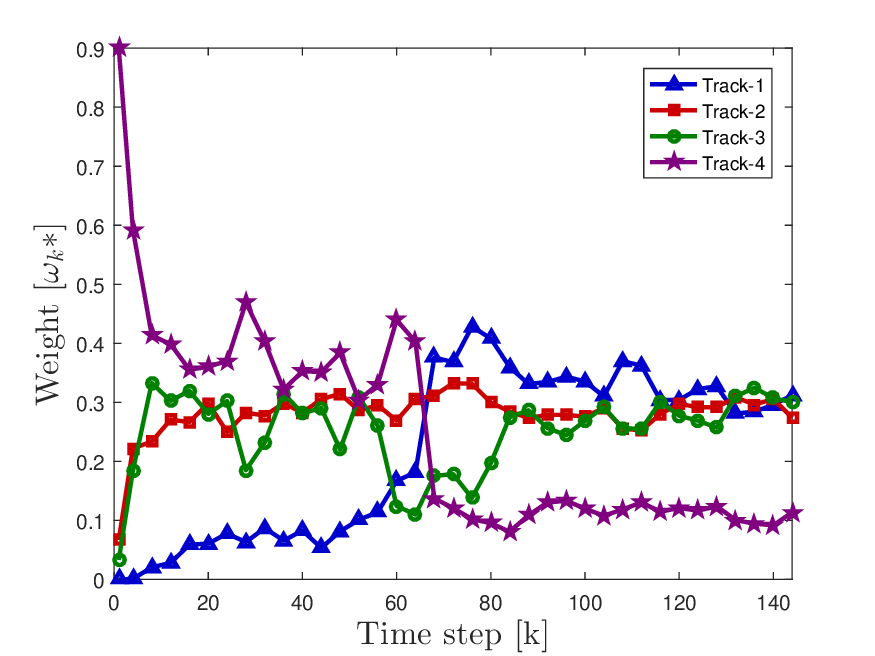}
	\caption{Weight variation of each track density over the time when target moves with a constant velocity (scenario~1).}
	\label{fig:opt_weight_CV}
   \end{minipage}
	\hfill
	\begin{minipage}[b]{0.48\linewidth}
		\centering
		\includegraphics[width=\linewidth]{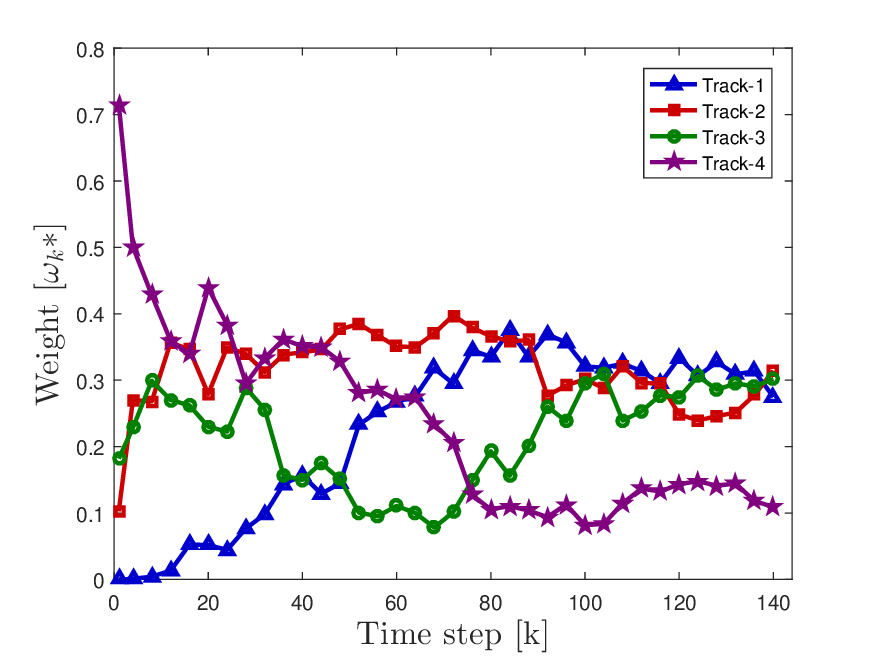}
		\caption{Weight variation of track densities over the time when target moves with a constant turn rate (scenario-2).}
		\label{fig:opt_weight_CT}
	\end{minipage}

\end{figure*}

\subsection{Scenario 1: Target moving with nearly constant velocity}
In this scenario, we have considered a straight line motion of target, where it is modeled as nearly constant-velocity motion. At each instant of time all the trackers produces its own track by processing the the available measurements. These generates tracks shared to all of their neighboring trackers for fusion. During the fusion, a cost function is minimized using nonlinear programming and the corresponding optimal weights for each tracker’s pdf are shown in Fig.~\ref{fig:opt_weight_CV}. From the figure, it could be seen the consensus fusion assigns higher weights to trackers that are closer to the target, initially favoring tracker-4 and eventually tracker-1, with the weights updating dynamically as the target moves.

\begin{figure*}[htbp]
	\centering
	\begin{minipage}[b]{0.48\linewidth}
		\centering
		\includegraphics[width=\linewidth]{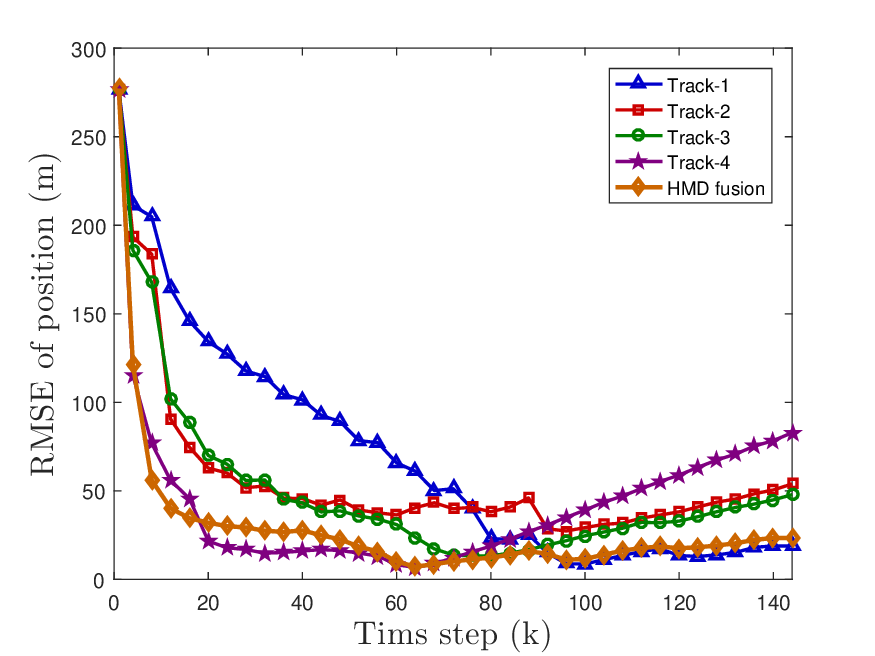}
		\caption{RMSE of position using SRF for each tracker and the  weighted HMD fusion when the target moves with constant velocity (scenario 1).}
		\label{fig:rmserc}
	\end{minipage}
	\hfill
	\begin{minipage}[b]{0.48\linewidth}
		\centering
		\includegraphics[width=\linewidth]{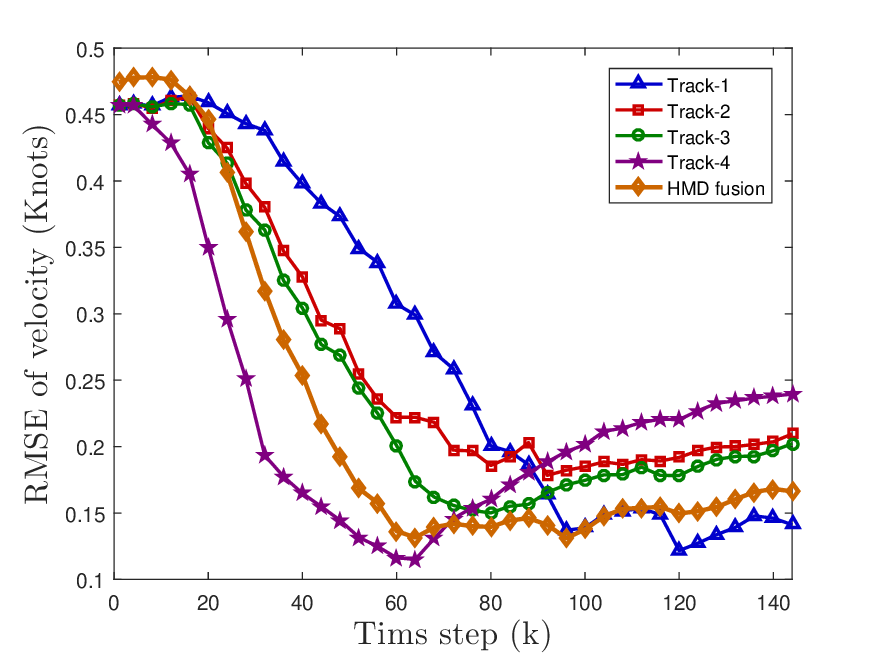}
		\caption{RMSE of velocity using SRF for each tracker and the  weighted HMD fusion when the target moves with constant velocity (scenario 1).}
		\label{fig:rmsevc}
	\end{minipage}
\end{figure*}

The RMSE results for both position and velocity are compared for all four trackers without fusion, alongside their  wighted harmonic mean density fusion using two selected sensors and the SRF, as shown in Figs. \ref{fig:rmserc} and \ref{fig:rmsevc}. The results indicate that track fusion consistently follows the most accurate trajectory among the individual trackers throughout the entire tracking interval and achieves faster convergence.

\begin{figure*}[htbp]
	\centering
	\begin{minipage}[b]{0.48\linewidth}
		\centering
		\includegraphics[width=\linewidth]{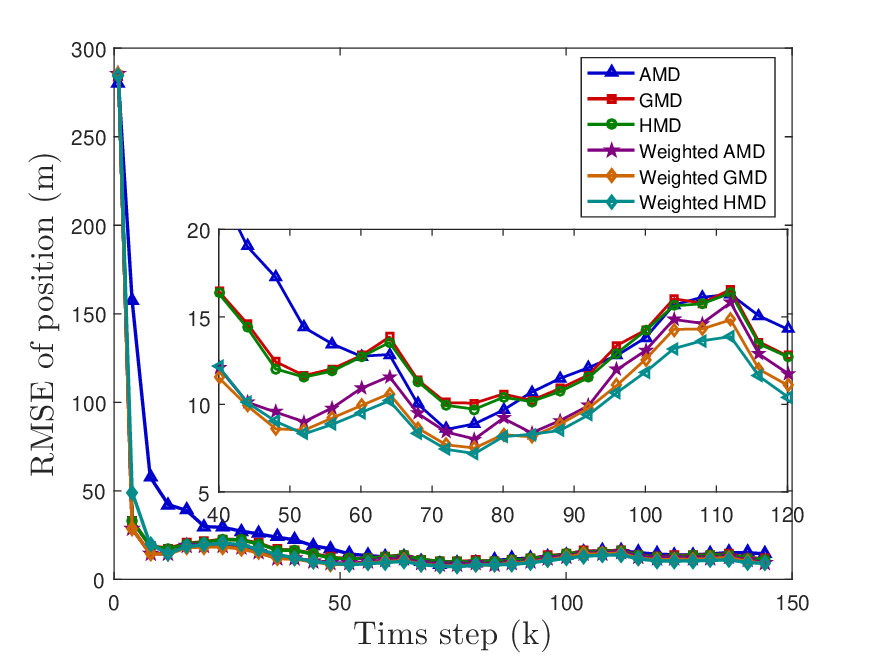}
		\caption{RMSE of position using SRF for different fusion methods using equal weights and optimized weights when the target moves with constant velocity (scenario~1).}
		\label{fig:rmserc3}
	\end{minipage}
	\hfill
	\begin{minipage}[b]{0.48\linewidth}
		\centering
		\includegraphics[width=\linewidth]{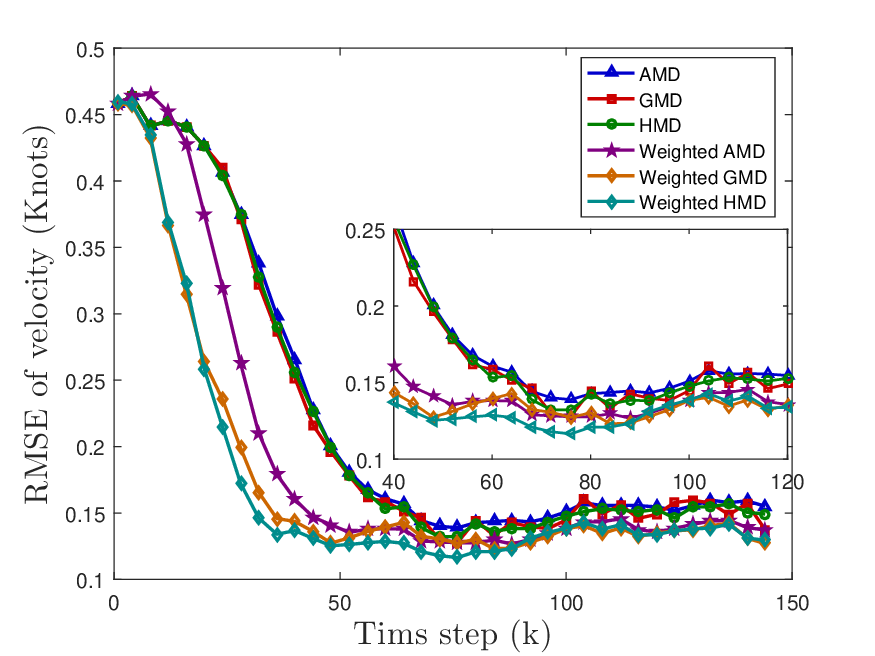}
		\caption{RMSE of velocity using SRF for different fusion methods using equal weights and optimized weights when the target moves with constant velocity (scenario~1).}
		\label{fig:rmsevc3}
	\end{minipage}
\end{figure*}

The RMSE results of both position and velocity are compared for consensus tracking by assigning equal weights and optimized weights for each tracker, where the optimized weights are obtained from the cost function using SRF, as shown in Figs.~\ref{fig:rmserc3} and \ref{fig:rmsevc3}. The results demonstrate that tracking performance with optimized weights is superior to that with equal weights, indicating that weight optimization is essential for achieving accurate consensus estimates. Furthermore, among the fusion methods with optimized weights, HMD and GMD exhibit similar performance.

\begin{figure*}[htbp]
	\centering
	\begin{minipage}[b]{0.48\linewidth}
		\centering
		\includegraphics[width=\linewidth]{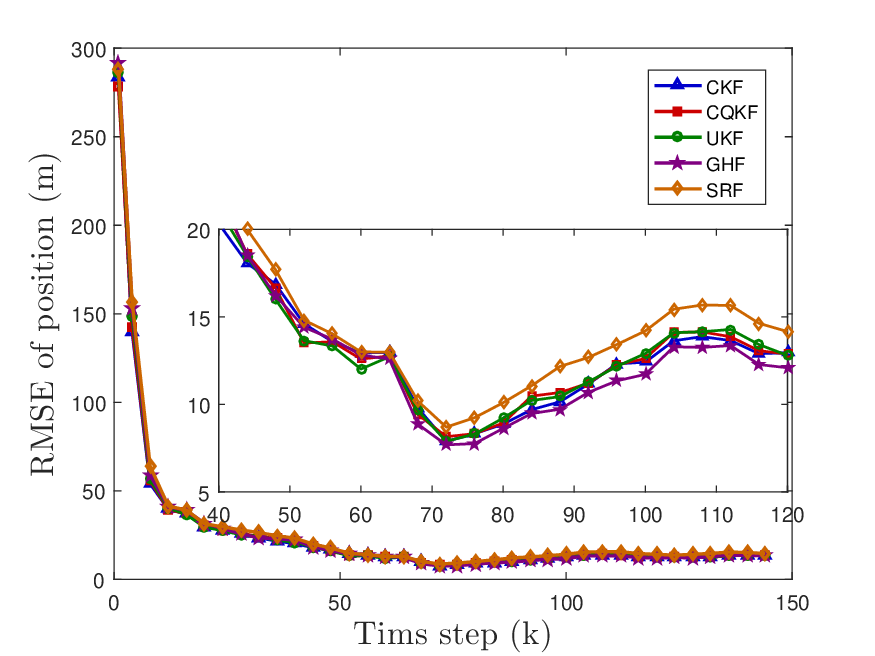}
		\caption{RMSE of position for  HMD fusion using different filtering techniques when the target moves with constant velocity (scenario~1).}
		\label{fig:rmserc4}
	\end{minipage}
	\hfill
	\begin{minipage}[b]{0.48\linewidth}
		\centering
		\includegraphics[width=\linewidth]{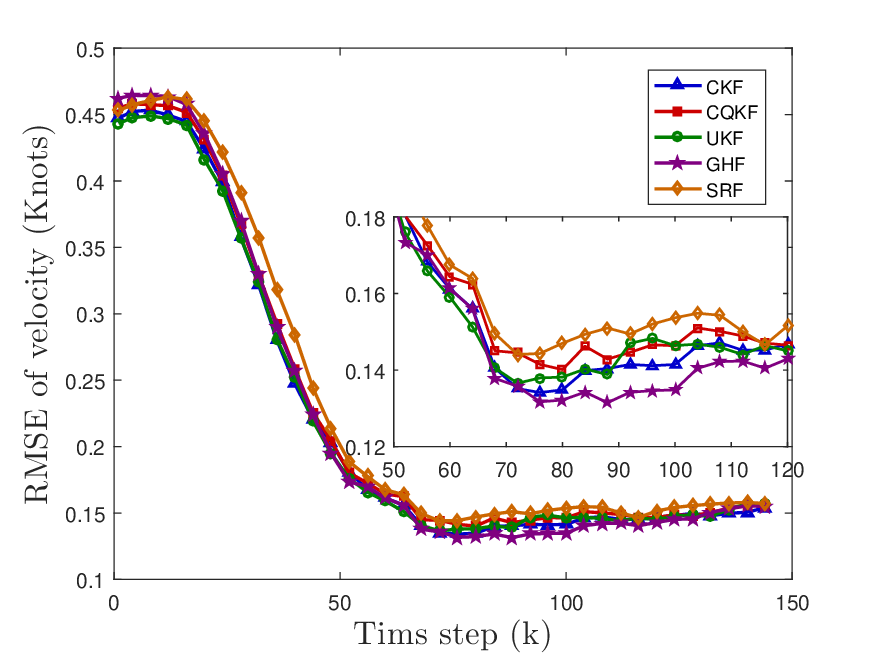}
		\caption{RMSE of velocity for  HMD fusion using different filtering techniques when the target moves with constant velocity (scenario~1).}
		\label{fig:rmsevc4}
	\end{minipage}
\end{figure*}

The RMSE results of both position and velocity are compared for  weighted harmonic density fusion based consensus tracking using different DSPFs \cite{bhaumik2019nonlinear} and SRF, as shown in Figs.~\ref{fig:rmserc4} and \ref{fig:rmsevc4}. From the figures, it can be seen that all the Gaussian filters designed for tracking perform well in consensus tracking, exhibiting almost identical performance.

\begin{figure*}[htbp]
	\centering
	\begin{minipage}[b]{0.48\linewidth}
		\centering
		\includegraphics[width=\linewidth]{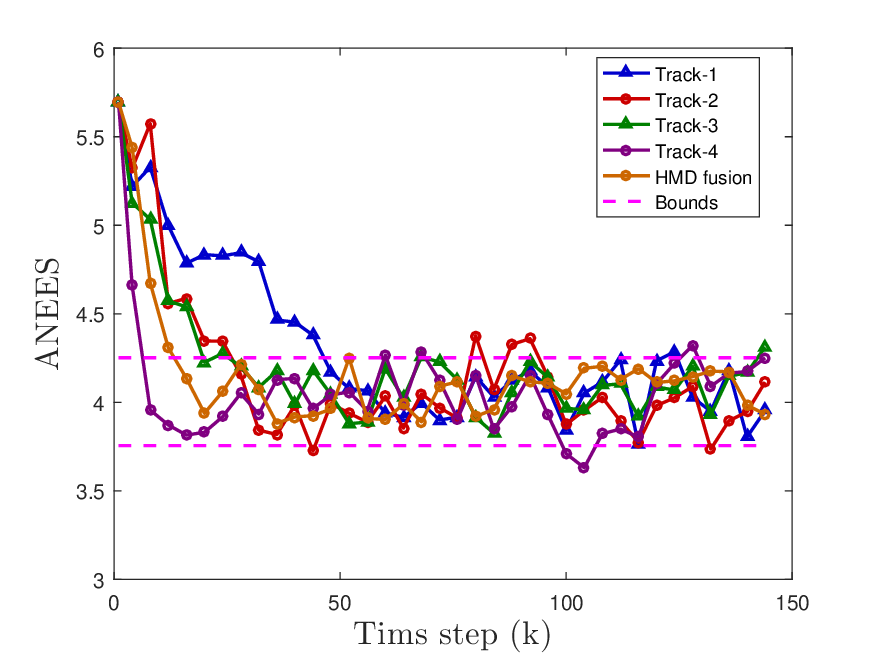}
		\caption{ANEES using SRF for each tracker and the  weighted HMD fusion when the target moves with constant velocity (scenario 1).}
		\label{fig:NEES_CV}
	\end{minipage}
	\hfill
	\begin{minipage}[b]{0.48\linewidth}
	\centering
	\includegraphics[width=\linewidth]{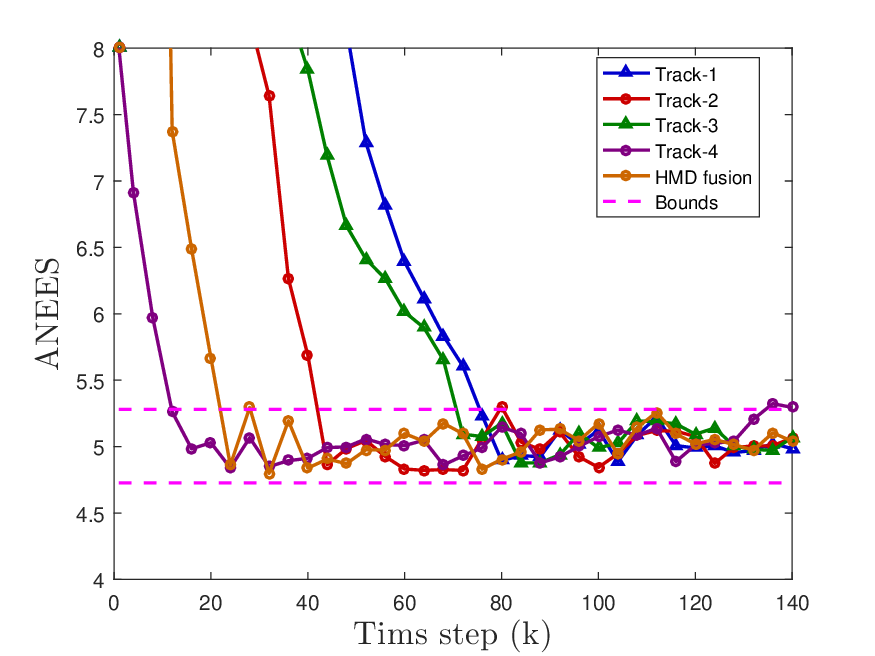}
	\caption{ANEES using SRF for each tracker and the  weighted HMD fusion when the target moves with constant turn rate (scenario 2).}
	\label{fig:NEES_CT}
\end{minipage}
\end{figure*}

The consistency of the proposed fusion is verified by the averaged normalized estimation error squared (ANEES), calculated using the relation:
\begin{equation*}
\text{ANEES}_k = \frac{1}{M_c} \sum_{m=1}^{M_c} 
(\mathcal{X}_{k}^m - \hat{\mathcal{X}}_{k|k}^m)^\top 
{P_{k|k}^m}^{-1} 
(\mathcal{X}_{k}^m - \hat{\mathcal{X}}_{k|k}^m),
\end{equation*}
where $M_c$ is the number of independent Monte Carlo simulations. The designed filter is considered consistent if, over time, the ANEES value lies within the 95\% confidence bounds $[l_b, u_b]$. These bounds for $n_{\mathcal{X}}$ states are defined as:
\begin{equation}
l_b = n_{\mathcal{X}} \left[ 
\Big(1 - \frac{2}{9 n_{\mathcal{X}} M_c} \Big) 
- 1.96 \sqrt{ \frac{2}{9 n_{\mathcal{X}} M_c} } 
\right]^3,
\end{equation}
\begin{equation}
u_b = n_{\mathcal{X}} \left[ 
\Big(1 - \frac{2}{9 n_{\mathcal{X}} M_c}\Big) 
+ 1.96 \sqrt{ \frac{2}{9 n_{\mathcal{X}} M_c} } 
\right]^3.
\end{equation}
The ANEES values of all four tracks without fusion are compared against those obtained using harmonic fusion tracking, as illustrated in Fig.~\ref{fig:NEES_CV}. The figure demonstrates that each individual sub-region tracker maintains consistent performance, with Tracker-4 reaching the 95\% confidence bounds earlier than the others. On the other hand, the consensus tracking achieved through harmonic fusion also converges to the confidence bounds and remains within the 95\% confidence region throughout the entire tracking period, indicating robust and reliable fusion behavior.

We also evaluated the ANEES performance for the other two fusion methods. Their comparison results are not reported here, but we found that the ANEES of the AMD-based fusion drops below the 95\% lower bound, indicating that the designed consensus filter with this method overestimates the uncertainty, even though its mean estimate remains closer to the true target. Instead, the other two methods perform well, with the advantage of HMD fusion being that it avoids non-integer powers of the track density, thereby simplifying implementation.

Table \ref{track_divergence} presents the percentage of track divergence for each sub-region’s track under different configurations in both scenarios. It compares the performance without fusion, with HMD fusion, and across different numbers of selected sensors, considering a track bound of $1.0$ km. The results show that as the number of sensors increases, the percentage of track divergence decreases. Moreover, applying HMD fusion among the distributed filters reduces the track divergence to zero. These findings demonstrate that the proposed HMD fusion technique effectively eliminates track divergence and enhances the robustness of the tracking system. Furthermore, increasing the number of sensors contributes to improved tracking accuracy and more consistent performance.
\begin{table*}[htbp]
	\caption{Percentage of track divergence for track bound = 1.0 Km}
	\begin{center}
		\begin{tabular}{ |c|c|p{1.4cm}|p{1.4cm}|p{1.4cm}|p{1.4cm}|p{1.4cm}| }
			\hline
			& Sensors & Track-1 & Track-2 & Track-3 & Track-4 & HMD \\
			\hline
			\hline
			Scenario 1 & $n = 2$ & 0 & 0.65 & 0.70 & 1.90 & 0 \\
			& $n = 3$ & 0 & 0.60 & 0.64 & 1.15 & 0 \\
			& $n = 4$ & 0 & 0.59 & 0.60 & 1.10 & 0 \\
			\hline
			Scenario 2 & $n = 2$ & 0.05 & 0.25 & 1.15 & 2.30 & 0 \\
			& $n = 3$ & 0 & 0.20 & 1.05 & 1.95 & 0 \\
			& $n = 4$ & 0 & 0.20 & 1.00 & 1.90 & 0 \\
			\hline
		\end{tabular}
		\label{track_divergence}
	\end{center}
\end{table*}

Performances of the various trackers are compared in terms of root mean square error (RMSE) averaged over the time span. It is termed as average root mean square error (ARMSE). ARMSE values for position (in meters) and velocity (in knots) across individual tracks without fusion, are compared to the results obtained using HMD fusion-based tracking in Table \ref{armse}. The comparison clearly demonstrates that the proposed fusion substantially improves tracking accuracy in both position and velocity.
\begin{table*}[htbp]
	\centering
	\caption{Average root mean square error (ARMSE) for position and velocity}
	\label{armse}
	\renewcommand{\arraystretch}{1.2}
	\resizebox{\textwidth}{!}{%
		\begin{tabular}{|c|c|cc|cc|cc|cc|cc|}
			\hline
			\multirow{2}{*}{} & \multirow{2}{*}{Sensors} 
			& \multicolumn{2}{c|}{Track-1} 
			& \multicolumn{2}{c|}{Track-2} 
			& \multicolumn{2}{c|}{Track-3} 
			& \multicolumn{2}{c|}{Track-4} 
			& \multicolumn{2}{c|}{HMD} \\ 
			\cline{3-12}
			& & Pos & Vel & Pos & Vel & Pos & Vel & Pos & Vel & Pos & Vel \\ 
			\hline
			\hline
			Scenario 1 & $n = 2$ & 67.81 & 0.2786 & 55.23 & 0.2660 & 47.19 & 0.2506 & 39.71 & 0.2228 & 22.29 & 0.2191 \\ 
			& $n = 3$ & 63.42 & 0.2684 & 53.94 & 0.2516 & 55.13 & 0.2408 & 37.14 & 0.2178 & 20.64 & 0.2118 \\ 
			& $n = 4$ & 61.98 & 0.2648 & 52.37 & 0.2438 & 44.56 & 0.2364 & 36.54 & 0.2142 & 19.36 & 0.2084 \\ 
			\hline
			Scenario 2 & $n = 2$ & 58.60 & 0.2786 & 39.80 & 0.2527 & 76.10 & 0.3402 & 43.40 & 0.2495 & 22.70 & 0.2333 \\ 
			& $n = 3$ & 55.40 & 0.2696 & 37.20 & 0.2484 & 71.50 & 0.3312 & 40.20 & 0.2378 & 21.10 & 0.2218 \\ 
			& $n = 4$ & 53.80 & 0.2658 & 36.40 & 0.2446 & 68.40 & 0.3272 & 38.70 & 0.2342 & 20.80 & 0.2164 \\ 
			\hline
		\end{tabular}%
	}
\end{table*}

\begin{figure*}[htbp]
	\centering
	\begin{minipage}[b]{0.48\linewidth}
		\centering
		\includegraphics[width=\linewidth]{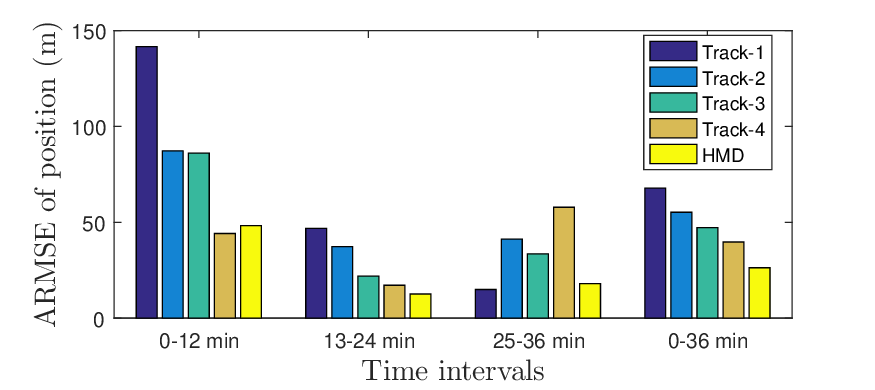}
		\caption{ARMSE variation over time using SRF for each tracker and the  weighted HMD fusion when the target moves with constant velocity (scenario~1).}
		\label{fig:ARMSE_bar_pos_CV}
	\end{minipage}
	\hfill
	\begin{minipage}[b]{0.48\linewidth}
		\centering
		\includegraphics[width=\linewidth]{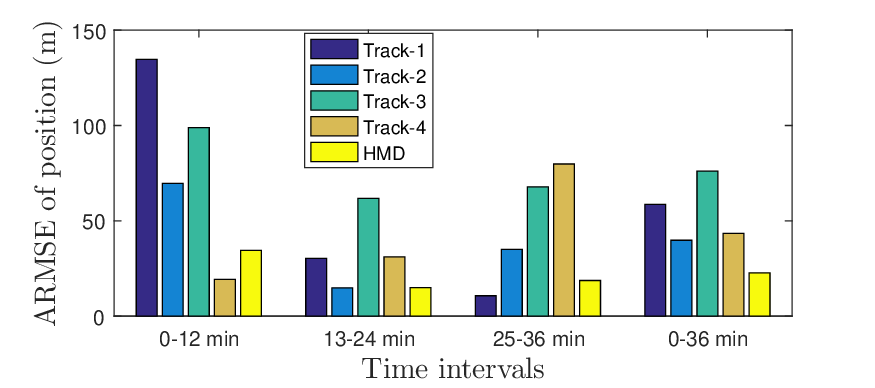}
		\caption{ARMSE variation over time using SRF for each tracker and the  weighted HMD fusion when the target moves with a constant turn rate (scenario~2).}
		\label{fig:ARMSE_bar_pos_CT}
	\end{minipage}
\end{figure*}

The ARMSE variation was computed by dividing the entire tracking interval into three 12-minute segments, and these segment-wise variations, along with the overall tracking period, are compared using the bar diagrams in Figs.~\ref{fig:ARMSE_bar_pos_CV} and \ref{fig:ARMSE_bar_pos_CT}. From these figures, it is clear that our fusion method consistently performs close to the best individual tracker, achieving the lowest overall ARMSE in both scenarios. Another observation is that the tracker closest to the target provides more accurate estimates, as measurements from nearby sensors exhibit larger variations, improving tracking performance. This relative performance changes over time as the target moves from one subregion to another.

\subsection{Scenario 2: Target moving with constant turn rate}
In this scenario, the target is modeled with a nearly coordinated turn motion, with a process noise intensity of $q_2 = 0.01 \; \text{rad}^2/\text{min}^3 $. All other parameters remain the same as in Scenario~1. The scenario is illustrated in Fig.~\ref{fig:scenario}, where the true target and its estimated trajectories are shown by dotted red and blue lines, respectively, with the coordinated turn clearly visible in the true path. The optimal weights for each tracker’s posterior densities  are shown in Fig.~\ref{fig:opt_weight_CT}. From the weight plot, it can be observed that the consensus fusion initially assigns higher weights to tracker-4, followed by tracker-2, and then tracker-1. This indicates that the fusion process gives higher weight to trackers that are closer to the target, relying more on their pdfs, with the weights dynamically changing over time as the target moves.

%
%

\begin{figure*}[htbp]
	\centering
	\begin{minipage}[b]{0.48\linewidth}
		\centering
		\includegraphics[width=\linewidth]{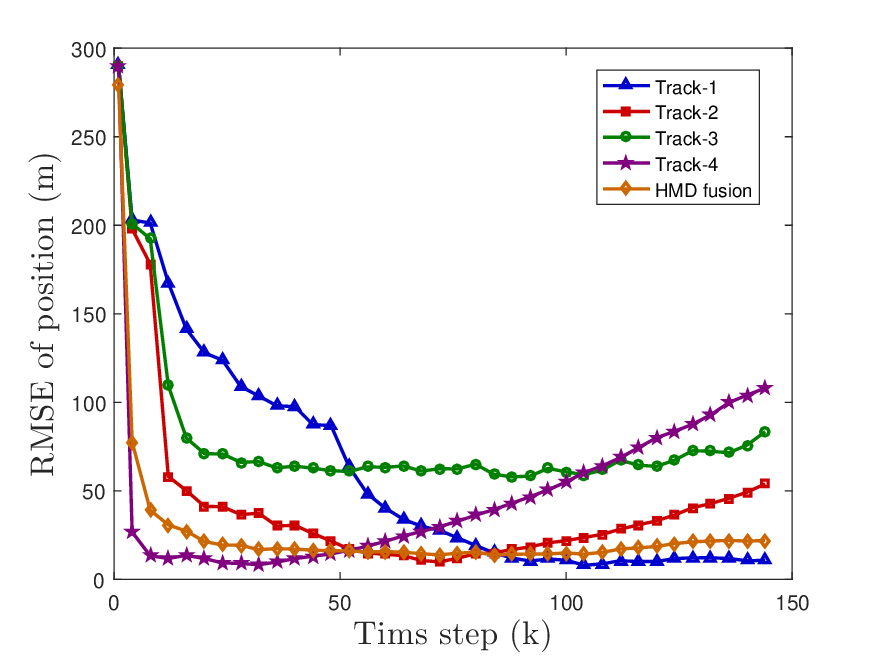}
		\caption{RMSE of position using SRF for each tracker and the  weighted HMD fusion when the target moves with constant turn rate (scenario 2).}
		\label{fig:rmserc_CT}
	\end{minipage}
	\hfill
	\begin{minipage}[b]{0.48\linewidth}
		\centering
		\includegraphics[width=\linewidth]{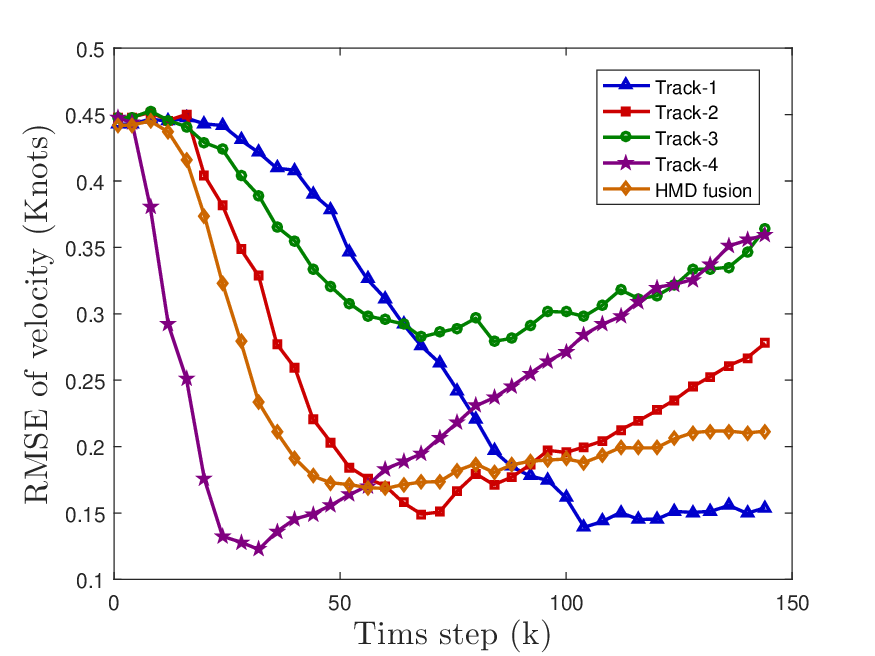}
		\caption{RMSE of velocity using SRF for each tracker and the  weighted HMD fusion when the target moves with constant turn rate (scenario 2).}
		\label{fig:rmsevc_CT}
	\end{minipage}
\end{figure*}

The RMSE results of both position and velocity are compared for all four trackers along with their HMD-fusion for two selected sensors and using SRF as shown in Figs. \ref{fig:rmserc_CT} and \ref{fig:rmsevc_CT}.  The figures show that track fusion-based estimation in consensus tracking always has follow the best performance of individuals in the entire tracking interval and provides faster convergence to all the trackers. 

\begin{figure*}[htbp]
	\centering
	\begin{minipage}[b]{0.48\linewidth}
		\centering
		\includegraphics[width=\linewidth]{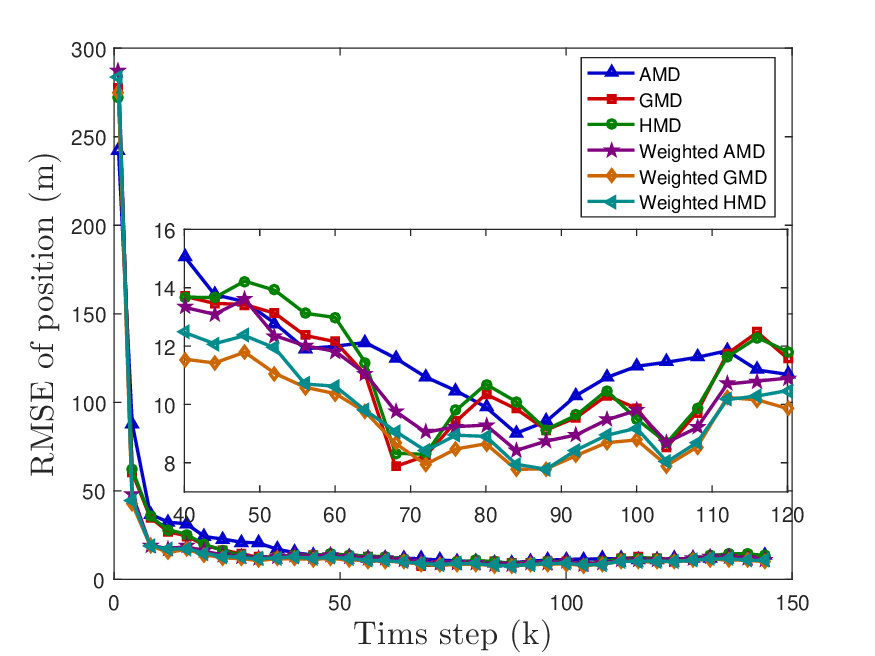}
		\caption{RMSE of position using SRF for different fusion methods using equal weights and optimized weights when the target moves with constant turn rate (scenario~2).}
		\label{fig:rmserc4_CT}
	\end{minipage}
	\hfill
	\begin{minipage}[b]{0.48\linewidth}
		\centering
		\includegraphics[width=\linewidth]{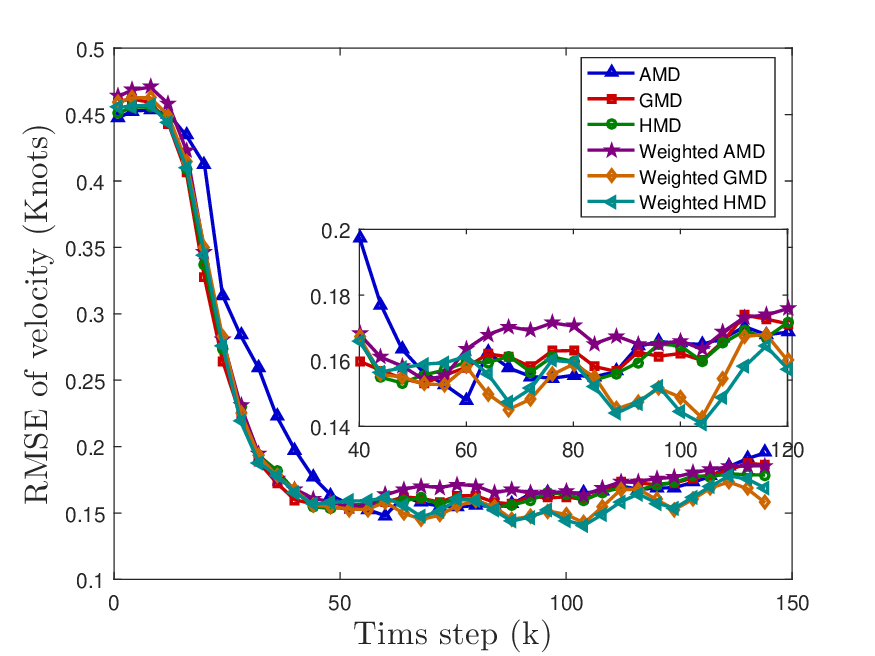}
		\caption{RMSE of velocity using SRF for different fusion methods using equal weights and optimized weights when the target moves with constant turn rate (scenario~2).}
		\label{fig:rmsevc4_CT}
	\end{minipage}
\end{figure*}

The RMSE results for both position and velocity are compared across different track density fusion methods in consensus tracking using SRF, considering both equal and optimized weights for each tracker. The optimized weights are obtained by minimizing the corresponding cost function, as illustrated in Figs.~\ref{fig:rmserc4_CT} and \ref{fig:rmsevc4_CT}. The results indicate that consensus tracking with optimized weights outperforms that with equal weights, highlighting the importance of weight optimization for accurate state estimation. Moreover, among the fusion methods employing optimized weights, the harmonic mean density (HMD) and geometric mean density (GMD) approaches exhibit nearly similar performance.

\begin{figure*}[htbp]
	\centering
	\begin{minipage}[b]{0.48\linewidth}
		\centering
		\includegraphics[width=\linewidth]{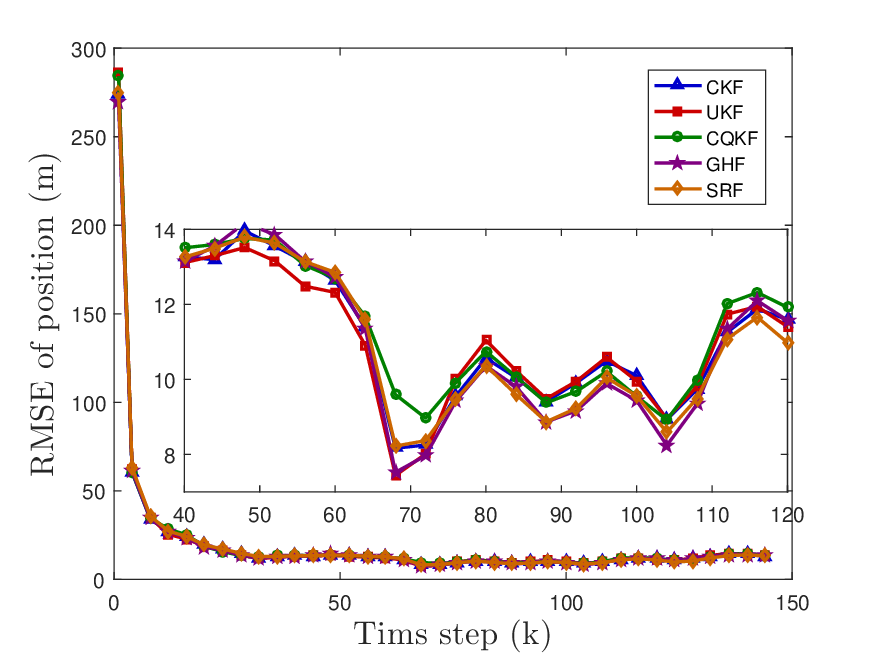}
		\caption{RMSE of position for  HMD fusion using different filtering techniques when the target moves with constant turn rate (scenario~2).}
		\label{fig:rmserc3_CT}
	\end{minipage}
	\hfill
	\begin{minipage}[b]{0.48\linewidth}
		\centering
		\includegraphics[width=\linewidth]{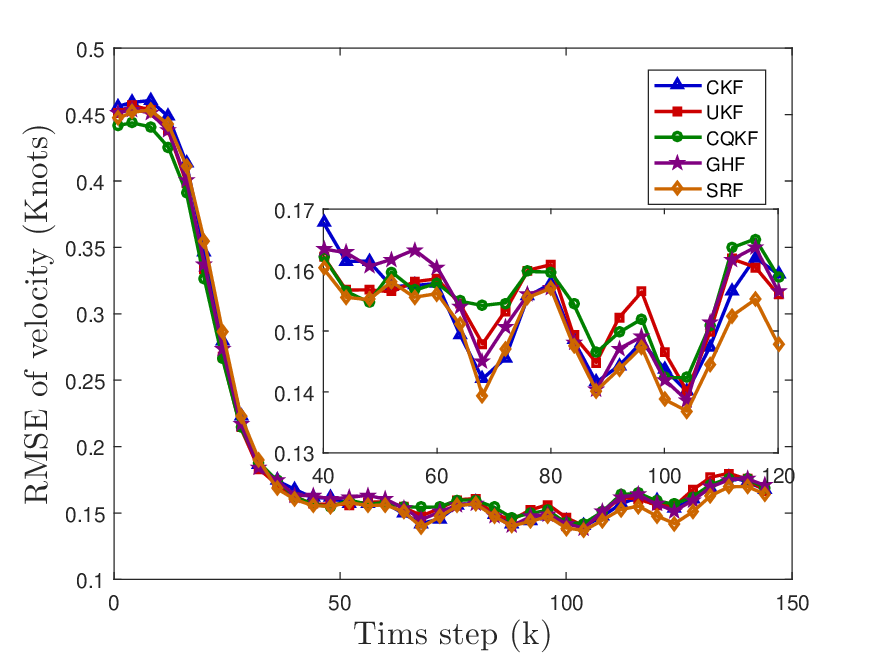}
		\caption{RMSE of velocity for  HMD fusion using different filtering techniques when the target moves with constant turn rate (scenario~2).}
		\label{fig:rmsevc3_CT}
	\end{minipage}
\end{figure*}

The RMSE results for both position and velocity are compared for  weighted consensus tracking using various deterministic sample point filters (DSPFs) \cite{bhaumik2019nonlinear} and the SRF, as shown in Figs.~\ref{fig:rmserc3_CT} and \ref{fig:rmsevc3_CT}. As illustrated, all the Gaussian filters designed for tracking perform effectively in the consensus framework, exhibiting nearly identical performance.

\appendices
\section{Derivation of the expression (29)}
\label{KL_for_Gaussian}
Let us consider two Gaussian pdfs, and can be expressed as 
\begin{equation*}
\begin{split}
&p_\ell(\mathcal{X}_k) = \\ &\tfrac{1}{(2\pi)^{\tfrac{n_\mathcal{X}}{2}}|{P_{k|k}^\ell}|^{\tfrac{1}{2}}}
\exp \big({-\tfrac{1}{2}(\mathcal{X}_k-\hat{\mathcal{X}}_{k|k}^\ell)^\top{P_{k|k}^\ell}^{-1}(\mathcal{X}_k-\hat{\mathcal{X}}_{k|k}^\ell)}\big),
\end{split}
\end{equation*}
where $ \ell \in \{i,j\} $. To compute the divergence, we compute
\begin{equation}
\begin{split}
\ln \tfrac{p_i(\mathcal{X}_k)}{p_j(\mathcal{X}_k)} =& 
\ln \tfrac{|{P_{k|k}^j}|^{\tfrac{1}{2}}}{|{P_{k|k}^i}|^{\tfrac{1}{2}}}
+ \tfrac{1}{2} (\mathcal{X}_k - \hat{\mathcal{X}}_{k|k}^j)^\top {P_{k|k}^j}^{-1} (\mathcal{X}_k - \hat{\mathcal{X}}_{k|k}^j) \\
&- \frac{1}{2} (\mathcal{X}_k - \hat{\mathcal{X}}_{k|k}^i)^\top {P_{k|k}^i}^{-1} (\mathcal{X}_k - \hat{\mathcal{X}}_{k|k}^i).
\end{split}
\label{eqk}
\end{equation}
By taking the expectation of the above expression over the first pdf \( \mathcal{X}_k  \sim \mathcal{N}(\mathcal{X}_k ; \hat{\mathcal{X}}_{k|k}^i, {P_{k|k}^i}) \)  is our desired KLD,  and it is computed by
\begin{equation}
\mathcal{ D}_{\mathrm{KL}}(p_i(\mathcal{X}_k) : p_j(\mathcal{X}_k)) = \mathbb{E}_{\mathcal{X} \sim p_i} \left[ \log \frac{p_i(\mathcal{X}_{k|k})}{p_j(\mathcal{X}_{k|k})} \right].
\end{equation}
The Eqn. (\ref{eqk}) contains the three terms; to avoid the complexity, let us compute the expectation of each term separately. The first term is independent of \( \mathcal{X} \), so expectation does not change its value, and it can be written as 

\begin{equation}
\mathbb{E} \bigg[\log \frac{|{P_{k|k}^j}|^{\tfrac{1}{2}}}{|{P_{k|k}^i}|^{\tfrac{1}{2}}}\bigg] = \frac{1}{2} \log \frac{|{P_{k|k}^j}|}{|{P_{k|k}^i}|}. \label{apxaeq1}
\end{equation}
The second term of Eqn.  (\ref{eqk}) is in quadratic form and by substituting $(\mathcal{X}_{k} - \hat{\mathcal{X}}_{k|k}^j) = (\mathcal{X}_{k} - \hat{\mathcal{X}}_{k|k}^i) + (\hat{\mathcal{X}}_{k|k}^i - \hat{\mathcal{X}}_{k|k}^j) $ it can be written as $  \big[ (\mathcal{X}_{k} - \hat{\mathcal{X}}_{k|k}^j)^\top {P_{k|k}^j}^{-1} (\mathcal{X}_{k} - \hat{\mathcal{X}}_{k|k}^j) \big]=  (\mathcal{X}_{k} - \hat{\mathcal{X}}_{k|k}^i)^\top {P_{k|k}^j}^{-1} (\mathcal{X}_{k} - \hat{\mathcal{X}}_{k|k}^i) + 2(\hat{\mathcal{X}}_{k|k}^i - \hat{\mathcal{X}}_{k|k}^j)^\top {P_{k|k}^j}^{-1} (\mathcal{X}_{k} - \hat{\mathcal{X}}_{k|k}^i) + (\hat{\mathcal{X}}_{k|k}^i - \hat{\mathcal{X}}_{k|k}^j)^\top {P_{k|k}^j}^{-1} (\hat{\mathcal{X}}_{k|k}^i - \hat{\mathcal{X}}_{k|k}^j),$ and its expectation is computed by
\begin{equation*}
\begin{split}
&  \mathbb{E}[(\mathcal{X}_{k} - \hat{\mathcal{X}}_{k|k}^i)^\top {P_{k|k}^j}^{-1} (\mathcal{X}_{k} - \hat{\mathcal{X}}_{k|k}^i)] = \text{tr}({P_{k|k}^j}^{-1} {P_{k|k}^i}),\\
&   \mathbb{E}[2 (\hat{\mathcal{X}}_{k|k}^i - \hat{\mathcal{X}}_{k|k}^j)^\top {P_{k|k}^j}^{-1} (\mathcal{X}_{k} - \hat{\mathcal{X}}_{k|k}^i)] = 0,\\
& \mathbb{E} [(\hat{\mathcal{X}}_{k|k}^i - \hat{\mathcal{X}}_{k|k}^j)^\top {P_{k|k}^j}^{-1} (\hat{\mathcal{X}}_{k|k}^i - \hat{\mathcal{X}}_{k|k}^j)] \\ &= (\hat{\mathcal{X}}_{k|k}^i - \hat{\mathcal{X}}_{k|k}^j)^\top {P_{k|k}^j}^{-1} (\hat{\mathcal{X}}_{k|k}^i - \hat{\mathcal{X}}_{k|k}^j).
\end{split}
\end{equation*}		

Thus, the expectation of the entire second term can be expressed as 
\begin{equation}
\begin{split}
& \mathbb{E}[\tfrac{1}{2}(\mathcal{X}_{k} - \hat{\mathcal{X}}_{k|k}^j)^\top {P_{k|k}^j}^{-1} (\mathcal{X}_{k} - \hat{\mathcal{X}}_{k|k}^j)]\\ = &   \tfrac{1}{2}(\hat{\mathcal{X}}_{k|k}^i - \hat{\mathcal{X}}_{k|k}^j)^\top {P_{k|k}^j}^{-1} (\hat{\mathcal{X}}_{k|k}^i - \hat{\mathcal{X}}_{k|k}^j) \\
& + \tfrac{1}{2}\text{tr}({P_{k|k}^j}^{-1} {P_{k|k}^i}) . \label{apxaeq2}
\end{split}
\end{equation}
The third term of Eqn.  \eqref{eqk} is also in quadratic form, and its expectation can be calculated by
\begin{equation}
\mathbb{E}[\tfrac{1}{2}(\mathcal{X}_{k} - \hat{\mathcal{X}}_{k|k}^i)^\top {P_{k|k}^i}^{-1} (\mathcal{X}_{k} - \hat{\mathcal{X}}_{k|k}^i)] = \text{tr}({P_{k|k}^i}^{-1} {P_{k|k}^i}) = \tfrac{n_{\mathcal{X}}}{2}. \label{apxaeq3}
\end{equation}
By combining all three terms from Eqns. (\ref{apxaeq1}), (\ref{apxaeq2}), and (\ref{apxaeq3}),  the KLD for two multivariate Gaussian pdfs can computed by
\begin{equation*}
\begin{split}
& \mathcal{ D}_{\mathrm{KL}}(p_i(\mathcal{X}_k ) : p_j(\mathcal{X}_k )) =
\tfrac{1}{2} \mathrm{tr}({{P_{k|k}^j}^{-1}} {{P_{k|k}^i}}) 	+ \tfrac{1}{2} \ln \frac{|{{P_{k|k}^j}}|}{|{{P_{k|k}^i}}|} \\
&   - \tfrac{ n_\mathcal{X}}{2} 
+ \tfrac{1}{2}({\hat{\mathcal{X}}_{k|k}^i} - {\hat{\mathcal{X}}_{k|k}^j})^\mathrm{T} {{P_{k|k}^j}^{-1}} ({\hat{\mathcal{X}}_{k|k}^i} - {\hat{\mathcal{X}}_{k|k}^j}),
\end{split}
\end{equation*}
where $n_\mathcal{X}$ is the dimension of the distributions and $|\cdot|$ denotes the determinant of a matrix.

\bibliographystyle{IEEEtran}
\bibliography{amef}

\end{document}